\documentclass[aps,prl,superscriptaddress,reprint]{revtex4-1}

\usepackage{graphicx}
\usepackage{graphics}
\usepackage{amsmath}
\usepackage[utf8]{inputenc}
\usepackage{color}
\usepackage{xcolor}
\usepackage{multirow}
\usepackage{todonotes}
\usepackage{fancyhdr}

\begin{document}

\title{Analysis of etching at a solid-solid interface}

%\title{Exact solution for the etching model}

\author{Washington S. Alves}
\email{wsalves@hotmail.com}
\affiliation{Graduate Program in Material Science, Faculdade UnB Planaltina, Universidade de Bras\'{i}lia, CEP 73300-000, Planaltina, DF, Brazil}

\affiliation{Instituto de F\'{i}sica, Universidade de Bras\'{i}lia, CP 04513, CEP 70919-970, Bras\'{i}lia, DF, Brazil}

\author{Evandro A. Rodrigues}

\affiliation{Instituto de F\'{i}sica, Universidade de Bras\'{i}lia, CP 04513, CEP 70919-970, Bras\'{i}lia, DF, Brazil}

\author{Henrique A. Fernandes}
\email{bernardo@fis.unb.br}

\affiliation{Universidade Federal de Goi\'{a}s, Campus Jata\'{i}, Br 364, Km 192, 3800, Parque Industrial, CEP 75801-615, Jata\'{i},
Goi\'{a}s, Brazil }

\author{Bernardo A. Mello}
\email{bernardo@fis.unb.br}

\affiliation{IBM Thomas J. Watson Research Center, Yorktown Heights, 10598 NY, USA  }

\author{Fernando A. Oliveira}
\email{faooliveira@gmail.com}
\affiliation{Instituto de F\'{i}sica, Universidade de Bras\'{i}lia, CP 04513, CEP 70919-970, Bras\'{i}lia, DF, Brazil}

\affiliation{Korea Institute for Advanced Study,  Seoul 130722, South Korea}

\author{Ismael V. L. Costa}
\email{ismaelcosta@gmail.com}
 
\affiliation{Graduate Program in Material Science, Faculdade UnB Planaltina, Universidade de Bras\'{i}lia, CEP 73300-000, Planaltina, DF, Brazil}

%\author{Ismael V. L. Costa\footnote{Corresponding author. Tel.: +55 61 31078002; fax: +55 61 3308 5185. E-mail address: %ismael@unb.br (Ismael Costa).}}

\begin{abstract}
We present a method to derive an analytical expression  for the roughness of an eroded  surface %of $1+1$ dimensions
whose dynamics are ruled by cellular automaton. Starting from the automaton, we obtain the time evolution of the height average and  height  variance (roughness).  
We apply this method to the etching model in $1+1$ dimensions, and then we obtain  the roughness  exponent. Using this in conjunction with the Galilean Invariance we obtain the other exponents, which perfectly match the numerical results obtained from simulations. These exponents are exact and they are the same as those exhibited by the Kardar-Parisi-Zhang (KPZ) model for this dimension.  Therefore, our results provide proof for the conjecture  that the etching and KPZ models belong to the same universality class. Moreover,  the method is general, and it can be applied to other cellular automata models.\\

Phys. Rev. E 94, 042119 – Published 17 October 2016

DOI: 10.1103/PhysRevE.94.042119

\end{abstract}


%\date{\today}

\maketitle

\section{Introduction}

%\paragraph*{Introduction} 
%{\bf
%Obtaining exact solutions is the main objective of most attempts to explain Nature from physical principles.  Nevertheless,  when interactions among many subunits are present most of the problems remain unsolved.   In statistical mechanics one can count  the models for  which the exact calculations of the critical exponents or dynamical exponents have been obtained. One of them is the famous  KPZ  growth model~\cite{Kardar86}  in $1 + 1$ dimensions.  We shall point out here that KPZ behaviour is very general, and arises even in some unexpected situations such as topological-defect turbulence in the electroconvection of nematic liquid crystals~\cite{Takeuchi10}. 
%We present a method to derive the analytical expression  of the roughness of an eroded  surface of $1+1$ 
% dimensions whose dynamics is ruled by cellular automaton. Starting from the automaton, we write down the time % % evolution for the height's average and  height's  variance (roughness).  
% We apply the method to the etching model~\cite{Bernardo}, then we obtain  the dynamical exponents, which % perfectly match the numerical results obtained from simulations. Those exponents are exact and they are the same %as those exhibited by the KPZ model for this dimension.  Therefore, they show that the etching model and KPZ %belong to the same universality class.}

%We present here the exact exponents for the etching model in $1 +1$ dimensions and we show that they are identical to those of  the KPZ model.

The dynamics of growth interfaces has become an interdisciplinary field and has had large success in the unification of concepts in many branches, ranging from applications in materials science~\cite{Yin05,Renner06} to fundamental problems in mathematics~\cite{Hairer13}. In the last decades two different approaches have been widely used. The first is the use of differential stochastic equations such as the KPZ equation~\cite{Kardar86,Hairer13,Sasamoto10}. 
However,  many recent advances are due to the availability of fast and affordable computer clusters. Such development has made it feasible, for instance, to obtain numerical simulations of large particle systems by obeying simple repetitive rules which mimic complex systems and whose emergent information and properties have been obtained through several techniques and approaches. 
The second method is defined as a cellular automata analysis.  It is obvious that there is a need to connect these methods, and considerable effort has been exerted to achieve this goal.
In this work we obtain the first exact solution for a growth cellular automaton model, i.e., we obtain the exact exponents for the etching model, and we show that it  belongs to the same class of universality as that of  the KPZ model.

The surface growth phenomenon, when treated as a stochastic process, encompass a
wide field of applications. Some examples of growth systems are corrosion~\cite{Mello01, Renner06,Reis03,Rodrigues15}, fire propagation~\cite{Merikoski03}, atomic
deposition~\cite{Csahok92}, evolution of bacterial colonies~\cite{Ben-Jacob94,
Matsushita90}, and spherical models~\cite{Henkel15}.
Models have
been proposed and studied through experiments~\cite{Ben-Jacob94,
Matsushita90}, analytical calculations~\cite{Barabasi95} and computational
simulations~\cite{Mello01,Reis03}.

Surfaces with different internal dynamics can lead to distinct profiles, which can
be characterized by different measures, the most important being the mean value
and the standard deviation of the surface height $h_i(t)$, $i=1,2...L$. When related to surfaces, the
standard deviation $w(t)$ is often called the surface width or roughness, and is defined as
\begin{equation}
w(t) =  \sqrt{\frac{1}{L} \sum_i {y_i}^2(t)},
\label{w0}
\end{equation}
wherein the variable $y_i$,  
\begin{equation}
y_i(t) = {h_i}(t) - \bar{h}(t),
\label{y}
\end{equation}
measures how high a site $i$ is  with respect to the mean substrate height
\begin{equation}
\bar{h}(t) = \frac{1}{L} \sum_i {h_i}(t).
\label{hm}
\end{equation}

Though the average height $\bar{h}(t)$ increases continuously due to the growth process, the
dynamic equilibrium leads to surface width saturation after a period of roughening
buildup. The saturated surface width, $w_s$, is often expresed as a function of the substrate size as the
power law $w_s\sim L^{\alpha}$, $\alpha$ being the roughness exponent. The
saturation occurs at times larger than a characteristic time $t_\times$ that follows the power
law $t_\times \sim L^{z}$, where $z$ is the dynamic exponent. Before saturation
($t\ll t_\times$), $w(L,t)$ evolves as a power law with the growth exponent
$\beta$, $w(L,t) \sim t^{\beta}$.

In the literature, we can find good examples of  cellular automata models~\cite{Katzav04,Predota96,Chua05,Buceta14} whose continuous versions are similar to the KPZ equation.  Although many works were dedicated at obtaining the exponents,
little effort was directed into obtaining a relation to describe the roughness time
 evolution.
 In this work we develop a general method to obtain the roughness evolution for cellular automata models.  Then we apply the method  to the etching model~\cite{Mello01}. We exactly obtain the  saturated roughness  $w_s$ and  the roughness exponent $\alpha$. From this we obtain $\beta$ and $z$, and we show that the etching model belongs to the same class of universality as that of the KPZ model.

\section{Basic theory}
\label{sec:Basictheory}

We develop here the basic ideas to obtain the roughness for a  cellular automaton model. The details are left to the appendix.  Consider the variation of roughness  squared for a given time interval  $\Delta t=1/L$ as
\begin{multline}
\frac{dw^2}{dt} =\lim_{\Delta t \rightarrow 0}\left<\frac{\Delta w^2}{\Delta t}\right>\\
= L\int P(w,y_{i-1},y_i,y_{i+1},t) \Delta w^2 dy_{i-1}dy_idy_{i+1}
%L\Phi(t)\int P_\text{eqp}(w,y_1,y_2,y_3) \Delta w^2 dy_1dy_2dy_3.\\
\label{wgeral0}
\end{multline}
 Here $\Delta w^2=\Delta w^2(y_{i-1},y_i,y_{i+1})$ is the variation in one possible event, i.e., we take a large number of experiments $N_e$  and at a time $t$ we take  a  variation $\Delta w^2$ that occurs in the next time interval.  This process is equivalent to being  weighted by the probability distribution $P(w,y_{i-1},y_i,y_{i+1},t)$, which depends on the configuration at the site $i$, its neighbors  $y_{i\pm1}$, and time. Note that values of $h_i$ are discrete variables. However, when we average over a large number of experiments all  possible real values of $y_i$ arise, and we can define the probability  $P$ and $\Delta w^2$  as continuous functions.

Unfortunately, there is no theory which can be used to obtain $P$.  In this way  we  define the auxiliary function 
\begin{equation}
\label{Phi0}
\Phi(t)=\frac{\int P(w,y_{i-1},y_i,y_{i+1},t) \Delta w^2 dy_{i-1}dy_idy_{i+1}}{\int P_\text{eqp}(w,y_{i-1},y_i,y_{i+1}) \Delta w^2 dy_{i-1}dy_idy_{i+1}},
\end{equation}
which we will call the {\em evolution factor}.  Here, $P_\text{eqp}(w,y_{i-1},y_i,y_{i+1})$ is an  equiprobability distribution. 
The integral on the denominator can be solved exactly.  With this definition we rewrite Eq.~(\ref{wgeral0}) as
\begin{equation}
\frac{dw^2}{dt}= L\Phi(t)\int P_\text{eqp}(w,y_{i-1},y_i,y_{i+1}) \Delta w^2 dy_{i-1}dy_idy_{i+1}
\label{wgeral007}
\end{equation}
We note that Eq.~(\ref{wgeral007}) is the same as Eq.~(\ref{wgeral0}), and no approximation were applied. 
The {\em evolution factor}  contains all information about the dynamics and correlation. It makes it possible to separate 
the time integral from the integral  in the configuration space. In this way the last  integral in  Eq.~(\ref{wgeral007}) can be exactly calculated. 
The factor $\Phi(t)$ will be  discussed later for the etching model.

\subsection{The equiprobability of the configurations}

To make the notation less
clumsy, the random  site will be chosen to be $i=2$, 
implying that only the sites $i=1, 2,3$ are relevant. Therefore, each step of the 
cellular automaton changes the  value of the squared surface width by $\Delta w^2(y_1,y_2,y_3)$. 

Before proposing an expression such as $P_\text{eqp}(w,y_1,y_2,y_3)$ we must remember
that there is a finite number of substrate configurations allowed by each value 
of $w$. This finite number is the result of the restrictions on $y_i$ imposed by
the definitions of $w^2$ and $y_i$ in Equations~(\ref{w0}) and~(\ref{y}),

\begin{subequations}
\label{conf}
\begin{align}
{y_1}^2+{y_2}^2+\dots+{y_L}^2 =& w^2L \label{confa}\\
y_1+y_2+\dots+y_L =& 0. \label{confb}
\end{align}
\end{subequations}
Eq.~(\ref{confa}) defines the surface of a hypersphere centered at the origin and of total radius $R_T=w\sqrt{L}$,
and Eq.~(\ref{confb}) defines a hyperplane, which passes at the origin, both in $L$-D, i.e., in a space of $L$ dimensions. From the
intersection of these two subspaces, a spherical surface results, which is
($L-2$)-D, and has the same radius $R_T$. 

For each combination of $y_1$, $y_2$, and $y_3$, i.e., fixed values of these variables, the
remaining $(L-3)$ $y_j$'s obeys
\begin{subequations}
\begin{align}
{y_4}^2+{y_5}^2+...+{y_L}^2 =& Lw^2-y_1^2-y_2^2-y_3^2=R^2 \\
y_4+y_5+...+y_L =& -(y_1+y_2+y_3), \label{restb}
\end{align}
\end{subequations}
and forms another spherical surface,
now with the dimensions $L-5$.
While the ($L-2$)-D surface corresponds to all possible surface configurations of a given value of $w$, the ($L-5$)-D surface is the subset of these configurations
defined by the triad $y_1$, $y_2$, and $y_3$. Following a common 
convention for the number of accessible states,
these surfaces will be called, respectively, $\Omega_T$ and $\Omega$. Consequently, the probability of a given triad is
\begin{equation}
P_\text{eqp}(w,y_1,y_2,y_3) = \frac{\Omega}{\Omega_T}. \label{pA}
\end{equation}

It is important to stress that the equiprobability assumption 
disregards the different probabilities of each configuration generated by
the cellular automaton, and even the fact that some configurations are never generated for certain automata. These selections rules are due to the cellular automaton and affect the function $\Delta w^2(y_1,y_2,y_3)$. However, one should always bear in mind that   all accessible states of a system, as defined by Eq.~(\ref{conf}), have equal probabilities as established by Statistical Mechanics.
A good test to check if the equiprobability distribution is suitable to 
describe the dynamics in a certain region is to observe if the behavior of $\Phi(t)$ is correct.  If it is constant in a given region, then equiprobability is present.

Eq.~(\ref{pA}) can be rewritten by substituting the expression for the area of a hypersphere,
\begin{equation}
P_\text{eqp}(w,y_1,y_2,y_3)= \frac{S_{L-5}\,{R}^{L-5}}{S_{L-2}\,{R_T}^{L-2}}
\label{p0}
\end{equation}
where $R$ and $R_T$ are the radius of the corresponding hyperspheres
and 
\begin{equation}
S_N = \frac{2 \pi^{(N+1)/2}}{\Gamma[(N+1)/2]},
\end{equation}
where $\Gamma(n+1)=n!$ is the gamma function. 

To obtain the radius $R$ we have to diagonalize the reduced hypersphereas shown in the Appendix.  The above probability can than be expressed more conveniently using the coordinates
\begin{equation}
\begin{array}{l}
y_1 = \sqrt{L}w\sin\rho\cos\theta\\
y_2 = \sqrt{L}w\sin\rho\sin
\theta\cos\varphi\\
y_3 = \sqrt{L}w\sin\rho\sin\theta\sin\varphi
\end{array}.\label{36c-1}
\end{equation}
with the variables defined in the intervals
\begin{equation}
0\leq\rho\leq\frac{\pi}{2},\quad 0\leq\theta\leq\pi,\quad 0\leq\varphi\leq2\pi.  \label{limits}
\end{equation}
After diagonalizing the small hypersphere (see Appendix), we arrive at 
\begin{multline}
\frac{dw^2}{dt}=L \Phi(t)\frac{S_{L-5}}{S_{L-2}} \int_0^{2\pi} \int_0^\pi \int_0^\frac{\pi}{2} 
\Delta w^2(w,\rho,\theta,\varphi)\,\sin^2\rho\,\\
\cos^{L-4}\rho\, \sin\theta\, d\rho\, d\theta\, d\varphi.
\label{wqfinal}
\end{multline}
Note that  Eq.~(\ref{DeltaW2}) and the above result are general, and consequently do not depend on of the cellular automaton used. However to obtain an explicit final result we need to select a model. In the next section we obtain this integral for the etching model

\section{The   etching model}
\label{sec:etchingmodel}

%\paragraph*{The etching model}

The etching model~\cite{Mello01} is an automaton that mimics the erosion of a surface by an acid. We randomly choose a site $i$, and we look at its  nearest neighbor. If $h_{i \pm 1} > h_i$, it is reduced to the same height as $h_i$, i.e., the height of the surface decreases at each step. We then defined it using the rules:\\

1. Randomly choose a site $i\in[1..L]$.

2. If $h_{i-1}(t)  >  h_i(t)$ do $h_{i-1}(t+\Delta t)=h_i(t)$.

3. If $h_{i+1}(t) >   h_i(t)$ do $h_{i+1}(t+\Delta t)=h_i(t)$.

4. Do $h_i(t+\Delta t)=h_i(t)-1.$\\

\begin{figure}
\centering
\includegraphics[scale=0.75]{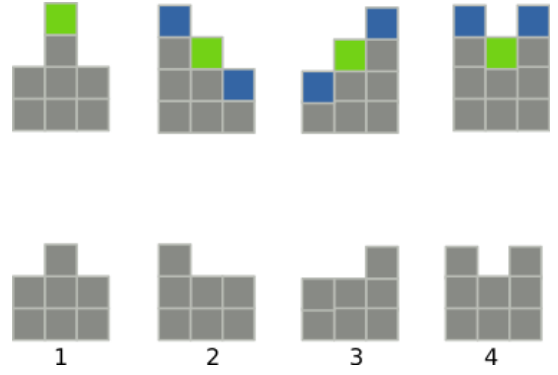}\\\vspace{3mm}
 \caption{The figures are the four situations in which the
 dynamics of the etching model can be divided. The 
 initial configuration is at the top, where the select size, column 2, is green and its neighbors are  shown in dark blue.
 After the corrosion process the final situation  is exhibited at the bottom.}
 \label{fig:neighbors}
\end{figure}

The algorithm implements a cell removal probability that is proportional to the
number of the exposed faces of the cell, a reasonable approximation of the
etching process. It could also describe a deposition where each exposed face
has the same attachment probability. For that reason, it can be referred to as either
particle removal, or deposition with $h \rightarrow -h$. The roughness is invariant to the   symmetry operation $h \rightarrow -h$. In Fig.~\ref{fig:neighbors} we show the evolution for the etching model.

%\onecolumngrid

\begin{table*}
\begin{tabular}{|c|c|ccc|c|c|}
\hline
 Case  & Limits & $\Delta y_1$ & $\Delta y_2$ & $\Delta y_3$ & $\sum\Delta y_i$ & $\sum(\Delta y_i)^2$  \\\hline
 1& $y_1<y_2>y_3$ & 0       &-1& 0       & -1           & 1 \\
 2& $y_1>y_2>y_3$ & $y_2-y_1$       &-1&    0    & $y_2-y_1-1$ &$1+{y_2}^2+{y_1}^2-2y_2y_1$\\
 3& $y_1<y_2<y_3$ & 0       &-1&$y_2-y_3$& $y_2-y_3-1$ &$1+{y_2}^2+{y_3}^2-2y_2y_3$\\
 4& $y_1>y_2<y_3$ & $y_2-y_1$&-1&$y_2-y_3$&$2y_2-y_1-y_3-1$&$1+{y_1}^2+2{y_2}^2+{y_3}^2-2y_2(y_1+y_3)$\\ \hline
\end{tabular}
 \caption{Change in some quantities that control the dynamics in each of the four cases
 	of Fig.~\ref{fig:neighbors}.}
 \label{tab:neighbors}
\end{table*}

Note that the change of the roughness has the term $(y_{2 \pm 1}-y_2)^2=(h_{2 \pm 1}-h_2)^2$ which is identical to the nonlinear term $(\nabla h)^2$ in the KPZ equation.
The scaling exponents found by Mello et al~\cite{Mello01} in $1+1$ dimensions were
$\alpha=0.491$ and $\beta=0.330$, suggesting that the model is within the KPZ
universality class  ($\alpha=1/2$, $\beta=1/3$) . Other studies
analyzed  the model in $1+1$ and $2+1$ dimensions focusing on aspects such as the
dynamic behavior of the roughness and comparisons with other models belonging
to the KPZ universality class~\cite{Mello15,Reis03, Reis04, Reis05, Ghaisas06,Oliveira08, Forgerini09}.
Some studies highlighted that the  Galilean invariance (GI) breaks down in fractal dimensions~\cite{Xun12},  
while recent work~\cite{Rodrigues15} shows that for $d+1$, $d$ integer and $\leq 6$, the  GI holds.  The last work also shows that inside this limit
there is no upper critical dimension. It is important to mention that there are some results showing that GI does not seem to play the
relevant role usually assumed in defining the KPZ universality class \cite{Wio10a,Wio10b}.
Finally some works demonstrate applications in the comprehension of kinetic roughening at metal-electrolyte interfaces~\cite{Cordoba07}.
Those works are  well-performed numerical analyses and are quite precise. However, we aspire to obtain exact solutions for this model. 
In the  appendix  we obtain $\Delta w^2 $ for the etching model. Using those results in Eq.~(\ref{wqfinal}),
we obtain  the  Eq.~(\ref{quadratic}) which may be rewritten as 
\begin{equation}
\label{dwt}
2w\frac{dw}{dt} = -c_a (w-w_+)(w-w_-)\Phi(t),
\end{equation}
where $w_\pm$ depends only on $L$. 
As shown in  Eq.~(\ref{ws}) for large values of $L$, $w_\pm = \pm w_s$ where
\begin{equation}
w_s \propto L^{1/2} .
\end{equation}
 This is the major result of our work  to obtain for the exact value for $\alpha$ for an automaton cellular model. For the etching model in $1+1$ dimensions it yields $\alpha=1/2$. We cannot at present time use our method to obtain $\beta$ or $z$. Furthermore, we cannot obtain the GI $z+\alpha=2$. However, numerical studies had been conducted in up to $6+1$ dimensions for the etching model~\cite{Rodrigues15}. If we  assume the GI, and  use the identity $z=\alpha/\beta$ we obtain  $\beta=1/3$ and $z=3/2$, which are the KPZ exponents. Consequently,  within the GI hypothesis,  we have obtained the exact exponents for the etching model,  and this adds to the evidence that it belongs to the same universality class as that of KPZ.
\begin{figure}
\centering
\includegraphics{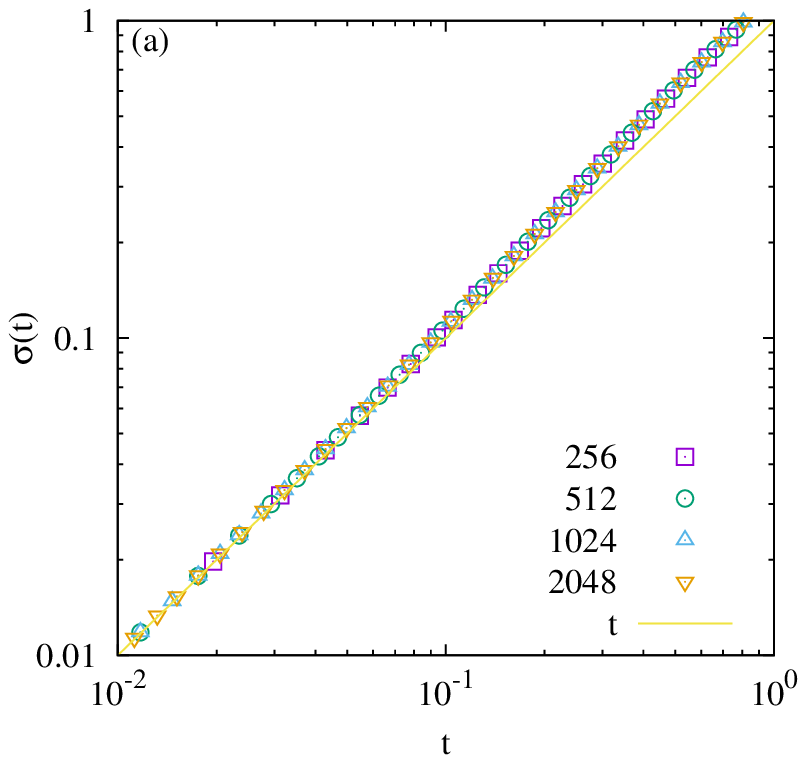}
\includegraphics{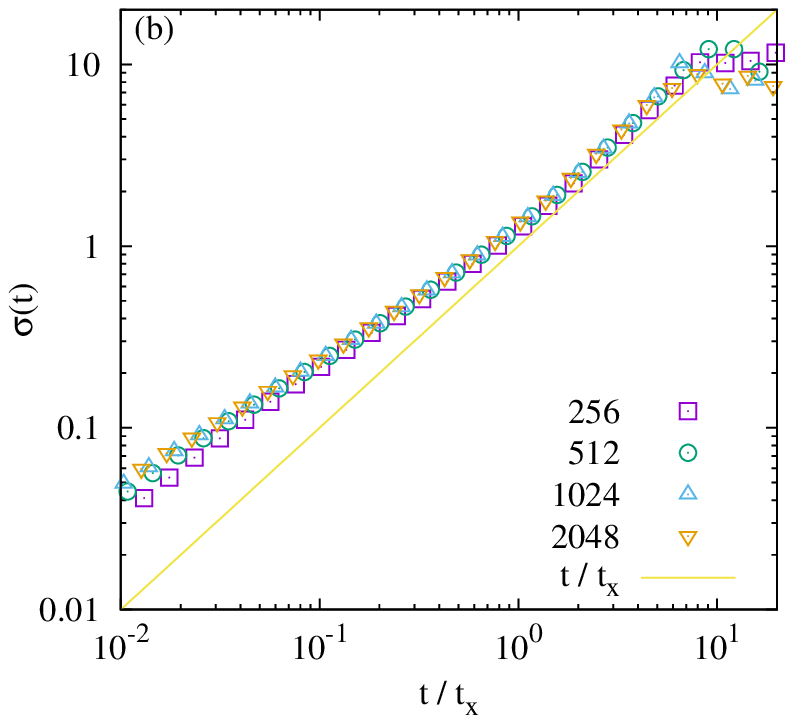}
\caption{\textbf{Evolution of} $\mathbf \sigma(t)$.
The data points are the results of numerical simulation of the etching model for several substrate lengths. The straight lines are the results for the uncorrelated process. (a) Expansion of $\sigma(t)$ for small values of $t$, implying small
values of $w(t)/w_s$. (b) Intermediate and large times.}
\label{fig:sigma}
\end{figure}

\section{The limiting behavior}

In the last section we  have achieved what we proposed for this work, i.e.\ to obtain 
the  exact exponents for the etching model for $1+1$ dimensions. Nevertheless, we feel 
that we need to better understand the time  evolution of $w(t)$.
The evolution factor $\Phi(t)$ contains all of the information about the dynamics and correlation
and remains to be determined.
We shall discuss it  in Fig.~\ref{fig:neighbors} and Tab.~\ref{tab:neighbors}.  
Now, we 
solve Eq.~(\ref{dwt}), with $w_\pm= \pm w_s$, to obtain
\begin{equation} 
\label{wt}
w(t) = w_s[1-\exp[(-\sigma(t)]]^{1/2}. 
\end{equation}
where
\begin{equation}
\sigma(t)=  \int_{0}^{t} \Phi(t') \, dt',
\end{equation}
 The  constant $c_a$ has been  absorbed in the above integral.
First note that we can reverse Eq.~(\ref{wt})  to obtain
\begin{equation}
\label{sigma}
\sigma(t)=-\ln{\left[1-[{w(t)}/w_s]^2\right]} . 
\end{equation}
Since $w(t)$  is a growing function of  $t$, $\sigma(t)$ is also a growing function of  $t$, which is in agreement with the fact that $\Phi(t)$ is always positive.

In Fig.~\ref{fig:sigma} we display  $\sigma(t)=-\ln{\left[1-[{w(t)}/w_s]^2\right]} $ as function of $t$. We obtain $w(t)$  by averaging over $10^7$ numerical experiments for several values of $L$. In Fig. 1a we show the beginning  of the growth process ($t \ll 1$, and $w(t) \ll w_s$), i.e.\ before the correlations become important, $\Phi(t)$ is a constant.  For that region   $\sigma(t) \propto w(t)^2 \approx t$, which is similar to uncorrelated random walk.   This is a general behavior for any growth process.
%The simulations shows it very clear on Fig 1a, where the data fit to a curve of the form  $\sigma(t) \propto t^b$, with $b=1.04$.  One can adjust as well within the same precision  $\sigma(t) =at$.  

In Fig.~\ref{fig:sigma}b we show $\sigma(t)$ for large $t$. Note that the time here is given in units of $t_\times$. For $1< t < t_\times$,  correlation becomes important, and $ w(t)  \propto t^{\beta}$, which means that $\sigma(t)$ no longer grows linearly with time. Finally, for $t >  t_\times$, $\sigma(t) \propto t$, i.e.,  we have left the region where the probability distribution is changing and arrived at
the region where that distribution is in the steady state, and, consequently, $\Phi(t)$ is constant. The straight line $\sigma(t)=\frac{t}{t_\times}$, i.e., the exact value for $\Phi(t)=1$  is displayed for comparison. We note that when $w(t)$ is very close to $w_s$,  the fluctuations make some values of $w(t)$ larger than $ w_s$ and the curve loses clarity and becomes blurred. If we increase the number of experiments, we decrease the uncertainty and the linear region increases. %\cite{Error}.

This shows the importance of the evolution factor $\Phi(t)$ and $\sigma(t)$ to describe the dynamics.  However, they are not necessary to obtain the exponent $\alpha$. This limiting behavior shows that in the beginning of the growth process, where no correlations are present, $\Phi(t)$ is constant. Likewise, for large times, when $w(t)$ approaches saturation, the configurations have established themselves and no longer change. Again, the equilibrium distribution perfectly describes the phenomenon.  This is identical to that which happens in anomalous diffusion~\cite{ Ferreira12}, where we do not have to know all of the stages of the evolution but we can obtain the exact exponents for the asymptotic  behavior $ t \rightarrow \infty$.   In this way we can obtain the exponents $\alpha$, $\beta$ and $z$ exactly, without a full knowledge of $\sigma(t)$.

\section{Conclusions}

In  this work, we present a method to obtain an equation for the roughness evolution. The definition of an evolution factor $\Phi(t)$ allows the separation of the time integral from  that of the configuration space.  This space is an $L$ dimensional space formed by the heights of the eroded surface.  The probability of occurrence of a given interface configuration is the ratio between the hypersphere area of that configuration
and the total hypersphere area in all possible configurations. In this way the statistical average in the configuration space can be obtained. 
The hypersphere method has potential for application in cellular automata
models which depend only on the nearest neighbors but needs to be built
differently for each type of algorithm. 
The only restriction on the algorithm 
is the condition that each site affects and is only acted on and affects the nearest neighbors.
A limitation of our method is that it provides exact results only at asymptotic times, neither
it provide the distribution of height $h(x,t)$, but the distribution of configurations. Nevertheless, it has the advantage of being a method for cellular automaton, an area which is mainly studied by numerical tools.

We applied the method to the $1+1$ etching model and obtained  the roughness exponent $\alpha$  which matches exactly with the KPZ exponent. If we consider that the GI is valid, which is supported by  numerical calculation~\cite{Rodrigues15},
 we can say that the etching model belongs to the KPZ universality class. 
 Recently, Sasamoto and  Spohn~\cite{Sasamoto10} have solved  the KPZ equation for $1+1$ dimensions. There, they found that the probability distribution function of the height $h(x,t)$ for all $t>0$. In particular, they  showed that  on the scale $t^{1/3}$, the statistics are given by the Tracy-Widom distribution~\cite{Tracy94,Fortin15}.  Moreover, we can say that we have obtained the stationary solution, $t> t_{\times}$, of the configuration distribution for the $1+1$ KPZ equation.  We hope that investigations of distribution evolution  in a lattice\cite{Oliveira10} can be used to obtain at lest one approximated  time dependent distribution of configurations.
 Finally,
 since our method does not involve renormalization approaches, which fails for 
$d \neq 1$,  the generalization for  $d>1$  is possible; nevertheless, it involves hard calculations.

\paragraph*{Acknowledgements} This work was supported by CAPES, CNPq, and FAPDF. FAO would like to thank the kind hospitality of Professor Hyunggyu Park  during his visit at the Korean Institute for Advanced Study (KIAS). 

\bibliography{corrosao}

\appendix

\section{Appendix}

\subsection{Evolution of the quadratic roughness}

The squared surface width will be affected by the value of ${y_i}^2$ before 
and after each step, for $i\in[1,3]$
\begin{equation}
\begin{split}
{y_i}^2(t) &=\left[h_i(t)-\bar{h}(t)\right]^2.\\
{y_i}^2(t+\Delta t) &= \left[h_i(t+\Delta t)-\bar{h}(t+\Delta t)\right]^2\\
  & = \left[{h_i}(t)+\Delta h_{i}-\bar{h}(t)-\Delta \bar{h}\right]^2\hspace{-0.7cm}\\
  & = \left[y_i(t) + \Delta h_i - \Delta \bar{h}\right]^2 \\
  & = y_i^2(t)+(\Delta y_i)^2
\end{split}
\end{equation}
with
\begin{equation}
\begin{split}
\Delta{y_i}^2 = & 2y_i(t)\Delta h_i - 2y_i(t)\Delta\bar{h}\\
& -2\Delta h_i\Delta\bar{h} + \left(\Delta {h_i}\right)^2 + \left(\Delta\bar{h}\right)^2.
\end{split}
\end{equation}
From Eq.~(\ref{confa}) we can write
\begin{equation}
\Delta w^2 = \frac{1}{L} \sum_i \Delta {y_i}^2. 
\end{equation}
A convenient expression for it can be found if we substitute 
the identities $\sum_i y_i=0$ and $\sum_i \Delta h_i = L\Delta\bar{h}$,
and remember that $\Delta h_i=0$ for $i \notin [1,3]$,
\begin{equation}
  \Delta w^2 = -\left(\Delta\bar{h}\right)^2+\frac{1}{L}\sum_{i=1}^3  \left[2y_{i}\Delta h_{i}+(\Delta h_{i})^2\right].
  \label{DeltaW2}
\end{equation}

Equation~(\ref{DeltaW2}) is a general formula (independent of the iterative algorithm) 
for the increment of the squared surface width at each time step, which
must be replaced in Eq.~(\ref{wgeral007}).
In order to obtain the dynamics of a specific algorithm, it is necessary to know the values of
$\Delta h_i$ and $\Delta\bar{h}$, which are functions of $w$, $y_1$, $y_2$, and $y_3$. 

\subsection{Diagonalizing the hypersphere}

The plane of Eq.~(\ref{confb}) contains the center of the sphere
of Eq.~(\ref{confa}), and therefore, the sphere defined by their intersection also has the same radius. However, the hyper plane~(\ref{restb}) does not cross the origin. Consequently we need to know the radius $R$. 
Squaring ~(\ref{restb}), we obtain
\begin{equation}
  -2\sum\limits _{\substack{i,j=4\\ i\neq j }}^{L}  y_i y_j =
   L w^2 - {y_1}^2 - {y_2}^2 - {y_3}^2-(y_1 + y_2 + y_3)^2,
\end{equation}
which we rewrite in a matrix  form:
\begin{multline}
 \begin{bmatrix} y_4 & y_5 & \cdots & y_L \end{bmatrix}
\begin{bmatrix}
    0 & -1 & \cdots & -1\\
   -1 & 0 & \ddots & \vdots\\
   \vdots & \ddots & \ddots & -1\\
   -1 & \cdots & -1 & 0
 \end{bmatrix}
  \begin{bmatrix} y_4 \\ y_5 \\ \vdots\\ y_L \end{bmatrix} =\\
  Lw^2 - {y_1}^2 - {y_2}^2 -{y_3}^2 - (y_1 + y_2 + y_3)^2.\label{matriz}
\end{multline}

The eigenvalues and a set of (non-orthogonal) eigenvectors of the above matrix are
\begin{align}
\lambda_4 =& 4-L\!\!\!\!& \lambda_5 &= 1 &\lambda_6&=1 \!\!&\cdots\;\;\; \lambda_L&=1\\
 v'_4 =& \begin{bmatrix}1\\1\\1\\1\\  \vdots\\1\end{bmatrix} \!\!\!\!& 
 v'_5 =& \begin{bmatrix}1\\-1\\0\\0\\ \vdots\\0\end{bmatrix} \!\!\!\!&
 v'_6 =& \begin{bmatrix}0\\1\\-1\\0\\ \vdots\\0\end{bmatrix} \!\!\!&\cdots\;
 v'_L =& \begin{bmatrix}0\\0\\ \vdots\\0\\1\\-1\end{bmatrix}
%\lambda_6 =& 1   & v'_6 =& (0,-1,1,0,\cdots,0)\\
%     \vdots& & \vdots\nonumber\\
%\lambda_L =& 1   & v'_L =& (0,\cdots,0,-1,1)
\end{align}
If these vectors are combined and rescaled to form an orthonormal basis,
$\{v_4,v_5,\cdots,v_L\}$, we
can define a linear isometric transformation to the set of variables ${y'_4, \dots,
y'_L}$,
\begin{equation}
\begin{bmatrix} y'_4  \\ y'_5  \\ \vdots \\ y'_L \end{bmatrix} =
\begin{bmatrix} v^T_4 \\ v^T_5 \\ \vdots \\ v^T_L \end{bmatrix}
\begin{bmatrix} y_4 \\ y_5 \\ \vdots\\ y_L \end{bmatrix} .
\end{equation} 
This transformation eliminates the off-diagonal terms of Eq.~(\ref{matriz}), 
\begin{multline}
 Lw^2 - {y_1}^2 - {y_2}^2 - {y_3}^2 - (y_1 + y_2 + y_3)^2 = \\
-(L-4){y'_4}^2 + {y'_5}^2 + \cdots + {y'_L}^2 ,
\end{multline}
 preserving the metrics used when calculating the
number of accessible states, $\Omega$ and $\Omega_T$.
The transformed variable $y'_4$ is equal to
\begin{equation}
 y'_{4} = \frac{1}{\sqrt{L-3}}(y_4+y_5+...+y_L).
\end{equation}
which, thanks to Eq.~(\ref{restb}) may be written as 
\begin{equation}
 y'_{4} = -\frac{1}{\sqrt{L-3}}(y_1+y_2+y_3) .
\end{equation}

With this transformation, Eq.~(\ref{matriz}) becomes
\begin{equation}
{y'_5}^2 + \cdots + {y'_L}^2 =
Lw^2 - {y_1}^2 - {y_2}^2 - {y_3}^2 -
\frac{(y_1 + y_2 + y_3)^2}{L-3}.
\end{equation}
The radius of this hypersphere is given by
\begin{equation}
 R =\sqrt{Lw^2-{y_1}^2-{y_2}^2-{y_3}^2-\frac{(y_1+y_2+y_3)^2}{L-3}}.
 \label{Rp2}
\end{equation}
We can now rewrite Eq.~(\ref{p0}) with the values $R_T$ and $R$ in the
asymptotic case $L\rightarrow\infty$,
\begin{equation}
P_\text{eqp}(w,y_1,y_2,y_3) = \frac{S_{L-5}}{S_{L-2}} \frac{\left[Lw^2 -{y_1}^2 -{y_2}^2  - {y_3}^2 \right]^\frac{L-5}{2}}{(Lw^2)^{\frac{L-2}{2}}}.
\label{p1}
\end{equation}

Now if we use coordinates~(\ref{36c-1}),
the probability~(\ref{p1}) expressed in these coordinates is
\begin{equation}
  P_\text{eqp}(w,\rho,\theta,\varphi) = 
   \frac{S_{L-5}}{S_{L-2}} \frac{1}{w^3 L^{3/2}} \cos^{L-5}\rho,
\end{equation}
and the Jacobian yields
\begin{equation}
dV=dy_1dy_2dy_3
 =w^3L^{3/2} \sin^2\rho \cos\rho \sin\theta\, d\rho\,d\theta\,d\varphi.
\end{equation}
Using the above results we arrive to Eq.~(\ref{wqfinal}).

\section{The etching model}

We will now calculate the evolution of the mean value of $w^2$
for the etching model. The value of $\Delta w^2$ in the
integrand of Eq.~(\ref{wqfinal}) is given by Eq.~(\ref{DeltaW2}), 
and therefore, we must determine  $\Delta h_i$ and $\Delta\bar h$
for each step of the etching model. To calculate these terms it is necessary to
separate the four possibilities of evolution of the etching model -- shown in Figure~\ref{fig:neighbors}
and Tab.~\ref{tab:neighbors}.

The values of Eq.~(\ref{DeltaW2}) corresponding
to the four cases shown on Tab.~\ref{tab:neighbors} are 
\begin{subequations}
\begin{align}
1:\enspace L\Delta w^2 = & 2y_2+1-\frac{1}{L} \\
2:\enspace L\Delta w^2 = & \left[1+(y_2-y_1)^2\right] \left(1-\frac{1}{L}\right)\\
      &+\frac{2}{L}(y_1-y_2)+2y_1(y_2-y_1)+2y_2\nonumber\\
3:\enspace L\Delta w^2 = & \left[1+(y_2-y_3)^2\right] \left(1-\frac{1}{L}\right)\\
    &+\frac{2}{L}(y_3-y_2)+2y_3(y_2-y_3)+2y_2\nonumber\\
4:\enspace L\Delta w^2 = & \left[1+(y_2-y_1)^2+(y_2-y_3)^2\right] \left(1-\frac{1}{L}\right)\\
      &+\frac{2}{L}(y_1+y_3-2y_2)-\frac{2}{L}(y_2-y_1)(y_2-y_3)\nonumber\\
      &+2y_1(y_2-y_1)+2y_3(y_2-y_3)+2y_2\nonumber
\end{align}
\end{subequations}
%https://www.sharelatex.com/project/54343f72e77e4f4d199354e8

The above expressions must be replaced in Eq.~(\ref{wqfinal})
with the limits of the table in Tab.~\ref{tab:neighbors}. The 
integrals are not complicated, but involve tedious calculations, resulting in 
\begin{equation}
  \frac{d w^2}{d t} =- [c_a w^2+ c_b w- c_c] \Phi(t),
  \label{quadratic}
\end{equation}
with
\begin{subequations}
\label{C}
\begin{align}
c_{a}&=\frac{1}{6}\frac{(7\pi-3\sqrt{3})(L-4)(L-2)L}{\pi(L-3)(L-1)^2}
\left(\frac{\Gamma\left(\frac{L+1}{2}\right)}{\Gamma\left( \frac{L}{2}+1\right)}\right)^2\\
c_{b}&=\frac{2\sqrt{2}(L-4)(L-2)\sqrt{L}}{\sqrt{\pi}(L-3)(L-1)^2}
\left(\frac{\Gamma\left(\frac{L+1}{2}\right)}{\Gamma\left( \frac{L}{2}+1\right)}\right)\\ 
c_{c}&=\frac{(L-4)(L-2)L}{2(L-3)(L-1)}
\left(\frac{\Gamma\left(\frac{L+1}{2}\right)}{\Gamma\left( \frac{L}{2}+1\right)}\right)^2 .
\end{align}
\end{subequations}

The polinomial of Eq.~(\ref{quadratic}) has the roots
\begin{equation}
  w_{\pm}=   \frac{-c_b \pm \sqrt{c_b^2+4c_ac_c}}{2c_a} .
  \label{roots}
\end{equation}
For large values of $L$ the coefficients of Eq.~(\ref{C}) behave as
$c_{c} /c_{a} \propto L $, and  $c_{b}/c_{a}  \propto L^0$, meaning that $w_\pm =\pm w_s$
with the asymptotic value of $w(t \rightarrow \infty)=w_s$,
\begin{equation}
w_s(L \rightarrow \infty)=\sqrt{\frac{c_c}{c_a}} \propto   L^{1/2}.  
    \label{ws}
\end{equation}

\end{document}